\documentclass[fleqn]{mn2e}
\usepackage{times}
\usepackage{amssymb}
\usepackage{amsmath}
\usepackage{graphicx}

\newcommand{\MBH}{M_{\text{BH}}}
\newcommand{\MDM}{M_{\text{DM}}}

\newcommand{\vc}{{v_{\text{c}}}}
\newcommand{\LB}{{L_{\text{B}}}}
\newcommand{\vcsigma}{$\vc$\,--\,$\sigma$}
\newcommand{\MBHsigma}{$\MBH$\,--\,$\sigma$}
\newcommand{\MBHvc}{$\MBH$\,--\,$\vc$}
\newcommand{\MBHLB}{$\MBH$\,--\,$\LB$}
\newcommand{\kms}{{\text{km\,s$^{-1}$}}}
\newcommand{\Msun}{{\text{M$_\odot$}}}
\newcommand{\vunit}{u_0}

\title[Supermassive black holes and dark matter haloes]%
{Observational evidence for a connection between supermassive black
holes and dark matter haloes}

\author[M. Baes et al.]%
{Maarten Baes$^{1,2}$\thanks{Postdoctoral Fellow of the Fund for
Scientific Research, Flanders, Belgium (F.W.O.-Vlaanderen)}, Pieter
Buyle$^1$, George K. T. Hau$^3$ and Herwig Dejonghe$^1$
\\$^1$Sterrenkundig Observatorium, Universiteit Gent, Krijgslaan 281
S9, B-9000 Gent, Belgium \\$^2$Department of Physics and Astronomy,
Cardiff University, 5 The Parade, Cardiff CF24\,3YB, Wales, UK
\\$^3$European Southern Observatory, Casilla 19001, Vitacura,
Santiago, Chile}

\begin{document}

\maketitle

\begin{abstract} 
We present new velocity dispersion measurements of sample of 12 spiral
galaxies for which extended rotation curves are available. These data
are used to refine a recently discovered correlation between the
circular velocity and the central velocity dispersion of spiral
galaxies. We find a slightly steeper slope for our larger sample, we
confirm the negligible intrinsic scatter on this correlation, and we
find a striking agreement with a corresponding relation for elliptical
galaxies. We combine this correlation with the well-known \MBHsigma\
relation to obtain a tight correlation between the circular velocities
of galaxies and the masses of the supermassive black holes they
host. This correlation is the observational evidence for an intimate
link between dark matter haloes and supermassive black holes. Apart
from being an important ingredient for theoretical models of galaxy
formation and evolution, the relation between $\MBH$ and circular
velocity can serve as a practical tool to estimate black hole masses
in spiral galaxies.
\end{abstract}


\section{Introduction}

Soon after the discovery of quasars in the 1960s, it was recognized
that only supermassive black holes (SMBHs) in the centres of galaxies
could provide the mechanism to feed active nuclei. An intriguing
problem arose however, when it became clear that quasars were much
more abundant in the early universe: what happened to these SMBHs~? A
natural explanation is that SMBHs still reside in the centres of
present day galaxies, but somehow the fueling of the active nucleus
stopped, for example due to the competition with star formation (Di
Mateo et al.~2003). The detection of SMBHs in quiescent galaxies is a
difficult task. Except for some special cases such as NGC\,4258
(Miyoshi et al.~1995) and the Milky Way (Ghez et al.~1998; Genzel et
al.~2000), the detection of a SMBH depends on stellar or gas
kinematics. In order to be able to actually detect the SMBH, it is
necessary to resolve the sphere of influence, i.e.\ the central region
where the black hole starts to dominate the potential of the
galaxy. With the limited resolution of ground-based spectrographs,
this was possible for only a handful of nearby galaxies. This changed
drastically with the advent of the {\em HST}, whose excellent spatial
resolution enabled to resolve the SMBH for a few dozen of nearby
galaxies (see Tremaine et al.~2002 and references therein). In nearly
all of the investigated galaxies, SMBHs were discovered with masses
roughly between 10$^7$ and 10$^9$ M$_\odot$. Very recently, the
detections of SMBHs of much lower masses have been reported in
globular clusters (Gebhardt, Rich \& Ho 2002; Gerssen et al.~2002,
2003), although these detections are still controversial (Baumgardt et
al.~2003a,b; Ho, Terashima \& Ulvestad~2003).

Having detected and measured SMBHs for a sizable sample of nearby
galaxies, we can proceed to tackle more fundamental questions
concerning their formation and evolution. A convenient way to do so is
by studying the relation between SMBHs and the galaxies that host
them. It was found that black hole masses are correlated with
parameters of the bulges of their host galaxies. Nearly ten year ago,
Kormendy \& Richstone~(1995) found a correlation between the mass
$\MBH$ of the SMBH and the blue magnitude $\LB$ of the bulge. More
recently, a tighter correlation between $\MBH$ and the velocity
dispersion $\sigma$ of the bulge was discovered independently by
Gebhardt et al.~(2000) and Ferrarese \& Merritt~(2000). Finally,
Graham et al.~(2001) presented an intriguing correlation between
$\MBH$ and the so-called concentration parameter of the bulge (or
S\'ersic index), in the sense that bulges with more massive black
holes have steeper cusps.

This apparently tight link between bulges and SMBHs reflects an
important ingredient that should be reproduced (and hopefully
explained) by theoretical models of galaxy formation. Actually, the
tightness of the correlations mentioned above is somewhat
surprising. Indeed, in most of the state-of-the-art models (e.g.\ Silk
\& Rees~1998; Adams et al.~2000; Kauffman \& Haehnelt~2000; Burkert \&
Silk~2001; MacMillan \& Hendriksen~2002; Wyithe \& Loeb~2002;
Volonteri, Haardt \& Madau~2003; Di Matteo et al.~2003), the total
galaxy mass (or dark matter mass $\MDM$), rather than the bulge mass,
plays a fundamental role in shaping the SMBHs. A close correlation
could therefore be expected between $\MBH$ and $\MDM$, rather than
between $\MBH$ and the properties of the bulge.

Unfortunately, this hypothesis is not straightforward to test
observationally, because the measurement of $\MDM$ and $\MBH$ are
difficult. An interesting approach to tackle this problem was taken by
Ferrarese~(2002). She argued that a correlation between $\MBH$ and
$\MDM$ should be reflected in a \MBHvc\ correlation, where $\vc$ is
the circular velocity in the flat part of the rotation curve of spiral
galaxies. Unfortunately, there are only four spiral galaxies with
secure SMBH masses (Genzel et al.~2000; Miyoshi et al.~1995; Bacon et
al.~2001; Lodato \& Bertin~2003), and only two of them have a measured
rotation curve. Therefore, Ferrarese used the tight
\MBHsigma\ correlation to estimate black hole masses in a larger
sample of galaxies: in fact, from a literature search, she found a
tight correlation between $\vc$ and $\sigma$ (with both $\vc$ and
$\sigma$ measured in \kms),
\begin{equation}
	\log\vc 
	= 
	(0.84\pm0.09) \log\sigma
	+
	(0.55\pm0.19).
\label{Ferrarese}
\end{equation}
The tightness of this correlation (it has a negligible intrinsic
scatter) suggests an intimate link between SMBHs and dark matter
haloes. A possible caveat in Ferrarese's analysis is that it is based
on only 16 spiral galaxies, including three spirals with
$\sigma<70~\kms$ which do not satisfy the
relation~(\ref{Ferrarese}). A larger database of galaxies with
extended rotation curves and velocity dispersion measurements would be
useful to check and refine this correlation. This is the goal of the
present Letter. We discuss the sample selection, the observations and
data reduction in Section~2. We analyze and discuss these data in
Section~3.

\section{The data set}

\begin{table}
\centering
\begin{tabular}{ccccc}
\hline \hline
name & type   & $\vc$  & $\sigma$ & $\log\MBH$ \\
     &        & (\kms) & (\kms)    & (\Msun)    \\            
\hline
\hline
NGC\,3038    & Sb   & 256\,$\pm$\,22 & 160\,$\pm$\,16 & 7.74\,$\pm$\,0.19 \\
NGC\,3223    & Sb   & 281\,$\pm$\,21 & 179\,$\pm$\,10 & 7.94\,$\pm$\,0.12 \\ 
NGC\,3333    & SBbc & 208\,$\pm$\,12 & 111\,$\pm$\,23 & 7.10\,$\pm$\,0.38 \\
ESO\,323-G25 & SBbc & 228\,$\pm$\,15 & 139\,$\pm$\,14 & 7.50\,$\pm$\,0.19 \\
ESO\,382-G58 & Sbc  & 315\,$\pm$\,20 & 165\,$\pm$\,22 & 7.80\,$\pm$\,0.24 \\
ESO\,383-G02 & SBc  & 190\,$\pm$\,14 & 109\,$\pm$\,28 & 7.07\,$\pm$\,0.46 \\
ESO\,383-G88 & SBc  & 177\,$\pm$\,16 &  70\,$\pm$\,14 & 6.30\,$\pm$\,0.38 \\
ESO\,445-G15 & Sbc  & 190\,$\pm$\,21 & 113\,$\pm$\,13 & 7.13\,$\pm$\,0.22 \\
ESO\,445-G81 & SBbc & 233\,$\pm$\,9  & 134\,$\pm$\,9  & 7.43\,$\pm$\,0.14 \\
ESO\,446-G01 & Sbc  & 213\,$\pm$\,17 & 123\,$\pm$\,12 & 7.28\,$\pm$\,0.19 \\
ESO\,446-G17 & Sbc  & 199\,$\pm$\,14 & 145\,$\pm$\,17 & 7.57\,$\pm$\,0.22 \\
ESO\,501-G68 & Sbc  & 173\,$\pm$\,9  & 100\,$\pm$\,16 & 6.92\,$\pm$\,0.30 \\ 
\hline
Milky Way    & SBbc & 180\,$\pm$\,20 & 100\,$\pm$\,20 & 6.92\,$\pm$\,0.37 \\
M31          & Sb   & 240\,$\pm$\,20 & 146\,$\pm$\,15 & 7.58\,$\pm$\,0.19 \\
M33	     & Sc   & 135\,$\pm$\,13 &  27\,$\pm$\,7  & 4.63\,$\pm$\,0.54 \\
M63          & Sbc  & 180\,$\pm$\,5  & 103\,$\pm$\,6  & 6.97\,$\pm$\,0.15 \\
NGC\,801     & Sc   & 216\,$\pm$\,9  & 144\,$\pm$\,27 & 7.56\,$\pm$\,0.34 \\
NGC\,2841    & Sb   & 281\,$\pm$\,10 & 179\,$\pm$\,12 & 7.94\,$\pm$\,0.13 \\
NGC\,2844    & Sa   & 171\,$\pm$\,10 & 113\,$\pm$\,12 & 7.13\,$\pm$\,0.21 \\
NGC\,2903    & SBbc & 180\,$\pm$\,4  & 106\,$\pm$\,13 & 7.02\,$\pm$\,0.24 \\
NGC\,2998    & SBc  & 198\,$\pm$\,5  & 113\,$\pm$\,30 & 7.13\,$\pm$\,0.48 \\
NGC\,3198    & SBc  & 150\,$\pm$\,3  &  69\,$\pm$\,13 & 6.27\,$\pm$\,0.36 \\
NGC\,4062    & SBc  & 154\,$\pm$\,13 &  90\,$\pm$\,7  & 6.73\,$\pm$\,0.19 \\
NGC\,4258    & SBbc & 210\,$\pm$\,20 & 138\,$\pm$\,18 & 7.48\,$\pm$\,0.24 \\
NGC\,4565    & Sb   & 264\,$\pm$\,8  & 151\,$\pm$\,13 & 7.64\,$\pm$\,0.17 \\
NGC\,5033    & Sc   & 195\,$\pm$\,5  & 122\,$\pm$\,9  & 7.27\,$\pm$\,0.16 \\
NGC\,6503    & Sc   & 116\,$\pm$\,2  &  48\,$\pm$\,10 & 5.64\,$\pm$\,0.42 \\
NGC\,7331    & Sbc  & 239\,$\pm$\,5  & 139\,$\pm$\,14 & 7.50\,$\pm$\,0.19 \\
\hline
\hline
\end{tabular}
\caption{
The galaxies in our sample. The galaxies above the horizontal line are
the galaxies from Palunas \& Williams~(2000) for which we have
obtained a velocity dispersion. The galaxies below the horizontal line
are the galaxies from Ferrarese~(2002) with a rotation curve extending
beyond the optical radius. The first two columns contain the name and
morphological type of the galaxies. The third and fourth columns
contain the circular velocity and bulge velocity dispersion of the
galaxies, with errors. The last column contains an estimate of the
SMBH mass, based on the \MBHsigma\ relation (see text).}
\end{table}

\begin{figure*}
\centering
\includegraphics[bb=48 307 550 534,width=0.9\textwidth,clip]{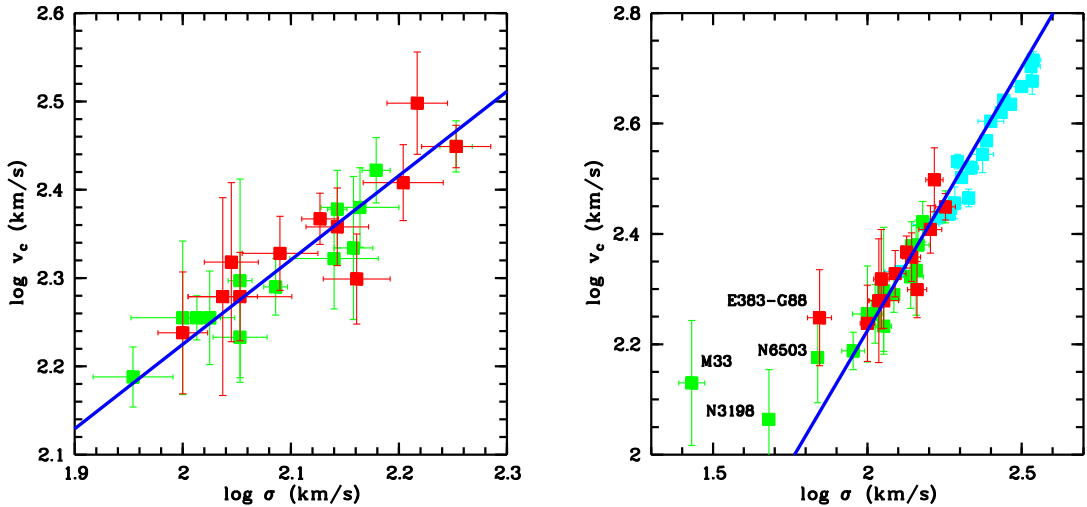}
\caption{
The correlation between the circular velocity $\vc$ and the central
velocity dispersion $\sigma$. The left plot shows the \vcsigma\
correlation for the 24 spiral galaxies with rotation curve beyond the
optical radius and a velocity dispersion $\sigma>80$~\kms. The green
dots are the data points from the sample of Ferrarese~(2002), the red
dots are the galaxies with new velocity dispersion from this
paper. The right plot zooms out and adds the elliptical galaxies from
Kronawitter et al.~(2000) to this plot. The straight line in both
panels represents equation~(\ref{vcsigma2}).}
\label{vcsigma.eps}
\end{figure*}

We aim to increase the number of galaxies with accurate measurements
of the circular velocity and the central velocity dispersion, in order
to investigate a possible link between SMBHs and dark haloes. It is a
well-known fact that the velocity dispersion of galaxies is a fairly
difficult quantity to measure accurately: values for the velocity
dispersion quoted in the literature often differ considerably
depending on the authors' favorite fitting techniques, template stars,
absorption line, signal-to-noise, etc. In order to construct a
reliable sample of galaxies, we therefore chose to select galaxies
from a homogeneous sample with measured extended rotation curves, and
to measure the velocity dispersions in a consistent way.

Our galaxies were drawn from the sample of Palunas \& Williams (2000),
who present a sample of 74 southern spiral galaxies. For each of these
galaxies they constructed axisymmetric mass models based on accurate
two-dimensional H$\alpha$ velocity fields and $I$-band imaging. 32 of
these galaxies have rotation curves extending to beyond the optical
radius\footnote{The optical radius used by Palunas \& Williams (2000)
is $R_{23.5}$, which is the radius of the isophote corresponding to an
{\em I}-band surface brightness of 23.5 mag\,arcsec$^{-2}$}, 26 of
which are reasonable smooth. These formed an ideal target for our
study. We were granted one night of observing time (4 May 2002) to
measure the velocity dispersion of these 26 galaxies at the ESO 3.6m
telescope in La Silla. Unfortunately, due to a very poor sky
transmission, the exposure times had to be multiplied by a factor 2.5
to obtain a sufficient signal-to-noise, such that we could only
observe 12 galaxies from the sample. The observed galaxies are listed
in Table~1.

We have taken long-slit spectra of these 12 galaxies with the EFOSC2
instrument through a 0.5 arcsec slit, giving a spectral resolution of
0.31 nm FWHM. We used the Gr\#08 grating, which enabled to measure the
velocity dispersions of the galaxies from the Mg{\sc i} and Fe lines
around 520~nm. The exposure time varied from 25 min to 150 min for the
faintest galaxies. Standard data reduction and calibration of the
spectra were done with the ESO-MIDAS package. This includes bias
subtraction and trimming, cosmics removal, wavelength calibration
(Legendre order 4 is used), removal of the S-distortion/tilt of the
slit, sky subtraction and airmass-correction. No effort was done to
flux-calibrate the spectra, as we are only interested in the velocity
dispersion. The reduced spectra have a wavelength domain between 430
and 630 nm, with a pixel scale of 9.8 nm/pixel.

The velocity dispersion $\sigma$ of the galaxies was determined by
means of a direct $\chi^2$ fitting technique, which fits line profiles
directly to the spectra after convolution with a stellar velocity
template spectrum. The fitting routine calculates the full covariance
matrix, from which we could determine the error bars on the
dispersion. The spatial aperture over which the fit was done was
chosen to maximize the signal-to-noise ratio of the spectrum, which
was targeted to be of the order of 15. In fact, our signal-to-noise
values ranged between 11 and 35. We found that the value of the
dispersion did not depend sensitively on the extent of the
aperture. The velocity dispersions of our 12 galaxies can be found in
the upper part of Table~1.

The circular velocities of galaxies were taken from Table~1 in Palunas
\& Williams~(2000). These are determined by taking a weighted
average of the rotation curve points where the rotation curve becomes
flat. We repeated this exercise on the original data, which were
kindly provided in tabular form by P. Palunas, to get an estimate for
the errors on the circular velocity. We find identical values as
Palunas \& Williams~(2000), with errors of the order of 5 to 10 per
cent.

\section{Analysis and discussion}

\begin{figure*}
\centering
\includegraphics[bb=48 307 550 534,width=0.9\textwidth,clip]{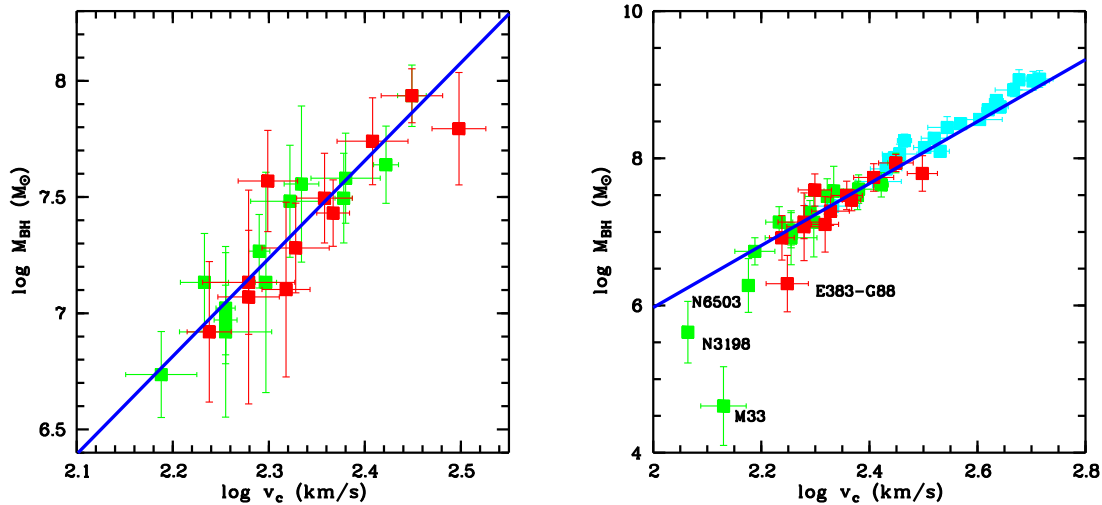}
\caption{
The correlation between the circular velocity $\vc$ and the SMBH mass
$\MBH$. Black hole masses are estimated from the velocity dispersions
through the \MBHsigma~relation, as given by
equation~(\ref{MBHsigma}). The layout is similar to
figure~{\ref{vcsigma.eps}}. The straight line in both panels
represents equation~(\ref{vcsigma2}).}
\label{MBHvc.eps}
\end{figure*}

\subsection{The correlation between $\vc$ and $\sigma$}

Combining our 12 galaxies with the 16 spiral galaxies from the
compilation of Ferrarese~(2002) with a rotation curve extending beyond
the optical radius (see Table~1), we obtain a total sample of 28
galaxies. In the left panel of figure~{\ref{vcsigma.eps}} we plot the
circular velocity versus the velocity dispersion for these 28
galaxies. For the 24 galaxies with a velocity dispersion greater than
about 80~\kms, there is a very tight correlation between $\vc$ and
$\sigma$. We fitted a straight line to these data, taking into account
the errors on both quantities (Press et al.~2002), and obtained
\begin{equation}
	\log\vc
	= 
	(0.96 \pm 0.11) \log\sigma
	+ 
	(0.32 \pm 0.25).
\label{vcsigma}
\end{equation}
In this correlation, both $\vc$ and $\sigma$ are expressed
in~\kms. However, in a linear regression analysis, this is not the
most meaningful unit to express these quantities. Tremaine et
al.~(2002) argued that the choice of appropriate units is important to
avoid a strong correlation of the errors in slope and zero-point. When
we repeat our fitting routine after choosing a new unit
$\vunit=200$~\kms, we obtain as a best fit
\begin{equation}
	\log\left(\frac{\vc}{\vunit}\right) 
	= 
	(0.96 \pm 0.11) \log\left(\frac{\sigma}{\vunit}\right) 
	+ 
	(0.21 \pm 0.023),
\label{vcsigma2}
\end{equation}
where indeed the uncertainty on the zero-point is much smaller.

Our newly derived \vcsigma\ relation has a slightly larger slope than
the original correlation~(\ref{Ferrarese}) found by Ferrarese~(2002)
based on 13 galaxies. Both slope and zero-point are consistent within
the one sigma error bar, though. The tightness of the correlation is
still astonishing: we find a reduced $\chi_r^2 = 0.281$, corresponding
to a goodness-of-fit of 99.9 per cent. The \vcsigma\ relation can
hence be regarded as having a negligible intrinsic scatter. Important,
moreover, is the fact that this correlation appears to be robust: even
with the sample of galaxies nearly doubled, there are still no
significant outliers. Ferrarese's \vcsigma\ correlation broke down for
galaxies with dispersions below about 80~\kms. We confirm this result:
the only galaxy we observed with $\sigma<80~\kms$ (ESO\,383-G88) joins
the three galaxies from Ferrarese's original sample (M33, NGC\,3198,
NGC\,6503 and ESO\,383-G88) in lying significantly above this
correlation.

It is interesting to compare this correlation to a corresponding one
recently found for elliptical galaxies. Based on stellar dynamical
models for 20 ellipticals constructed by Kronawitter et al.~(2000),
Gerhard et al.~(2001) discovered a very tight relation between the
central dispersion and the circular velocity (the circular velocity
curves of ellipticals were found to be flat to within 10 per
cent). Ferrarese~(2002) reports for these 20 ellipticals the
correlation
\begin{equation}
	\log \vc
	=
	(0.94 \pm 0.11) \log \sigma
	+
	(0.31 \pm 0.26),
\end{equation}
with a reduced $\chi_r^2 = 0.66$. Both the slope and zero-point of
this correlation agree amazingly well with the \vcsigma\
correlation~(\ref{vcsigma}) of our spiral galaxy sample. These
expressions represents a nearly direct proportionality between the
bulge velocity dispersion and the halo circular velocity, with the
proportionality constant about two thirds. To first order, elliptical
galaxies form a dynamically uniform family, such that we expect a
proportionality between $\sigma$ and $\vc$ (Gerhard et al.~2001). For
spiral galaxies, however, this proportionality cannot be explained by
simple dynamical arguments, as convincingly argued by
Ferrarese~(2002). The fact that there is such a strong correlation
between both velocity scales in spiral galaxies is hence not obvious,
and indicates a fundamental correlation in the structure of
spirals. Moreover, the fact that both spiral and elliptical galaxies
obey exactly the same correlation is absolutely striking.

\subsection{The correlation between $\MBH$ and $\vc$}

Although derived from mainly late-type spirals (all galaxies in our
sample except NGC\,2844 have a Hubble type of Sb or later) with
$\sigma$ between 90 and 180~\kms, the \vcsigma\ correlation appears to
hold as well for a larger class of galaxies, ranging from ellipticals
to late-type spirals, with a dispersion range up to 350~\kms. A
similar situation appears to apply for the
\MBHsigma\ relation. From the three recently discovered empirical
relations linking $\MBH$ to the bulge properties of the host galaxies
(see Introduction), the \MBHsigma\ relation is the tightest one. The
two original papers describing this correlation found significantly
different slopes: 4.8\,$\pm$\,0.5 (Ferrarese \& Merritt~2000) versus
3.75\,$\pm$\,0.3 (Gebhardt et al.~2000). There has been a vivid
discussion since over the exact slope of this relation, and it is
argued that these differences can be ascribed to different fitting
techniques, different samples and different measures of {\em the}
central velocity dispersion (Merritt \& Ferrarese~2001a, b; Tremaine et
al.~2002). The most up-to-date version of the \MBHsigma\ correlation,
from Tremaine et al.~(2002), reads
\begin{equation}
	\log\left(\frac{\MBH}{\Msun}\right)
	=
	(4.02 \pm 0.32) \log\left(\frac{\sigma}{\vunit}\right)
	+
	(8.13 \pm 0.06)
\label{MBHsigma}
\end{equation}
This relation is derived from empirical data for some 30 nearby
galaxies galaxies with secure SMBH detections. Most of these galaxies
are elliptical galaxies with only a minor fraction of spiral galaxies
present. Nevertheless, this correlation seems to hold for lenticular
and spiral galaxies with similar scatter, contrary to the \MBHLB\
correlation (McLure \& Dunlop~2002). Moreover, the SMBHs recently
claimed to be detected in the globular clusters M15 and G1, also
appear to satisfy the same \MBHsigma\ relation (Gebhardt et al.~2002).

As both the \vcsigma\ and \MBHsigma\ correlations seem to hold over
the entire Hubble range (see e.g.\ de Zeeuw~2003 for a critical note),
we can combine them to derive a correlation between the circular
velocity and SMBH mass. For all galaxies in our sample, we have
estimated $\MBH$ via the \MBHsigma\ correlation given by
equation~(\ref{MBHsigma}). The results can be found in table~1.
Combining our best fitting \vcsigma\ relation~(\ref{vcsigma2}) with
the \MBHsigma\ relation~(\ref{MBHsigma}), we obtain
\begin{equation}
	\log\left(\frac{\MBH}{\Msun}\right)
	=
	(4.21 \pm 0.60) \log\left(\frac{\vc}{\vunit}\right)
	+
	(7.24 \pm 0.17).
\label{MBHvc}
\end{equation}
In Figure~{\ref{MBHvc.eps}} (left panel) we plot the circular velocity
versus the SMBH mass for the 24 galaxies of our sample with
$\sigma>80~\kms$: we find indeed a very tight correlation between
$\MBH$ and $\vc$. Moreover, this correlation also holds for the
elliptical galaxies of Kronawitter et al.~(2000), as shown in the
right panel of Figure~{\ref{MBHvc.eps}}. For the least massive
galaxies with $\sigma<80~\kms$, this correlation again breaks down.

The correlation between SMBH mass and circular velocity is useful for
two different goals. Firstly, if a prescription can be found to link
the circular velocity of a galaxy to the total dark halo mass $\MDM$,
we can transform this \MBHvc\ correlation to a direct correlation
between SMBH mass and total galaxy mass, which can then be compared
with theoretical models of galaxy formation. A conversion between
these two quantities can in principle be derived from high-resolution
CDM cosmological simulations (e.g.\ Navarro \& Steinmetz~2000; Bullock
et al.~2001). Using the simulation results of Bullock et al.~(2001),
\begin{equation}
	\frac{\MDM}{10^{12}~\Msun}
	\sim
	1.40\left(\frac{\vc}{\vunit}\right)^{3.32},
\end{equation}
Ferrarese~(2002) convert $\vc$ to $\MDM$, and derives a relation
between SMBH and dark halo masses. If we repeat the same exercise with
our larger sample of spirals and the most up-to-date 
\MBHsigma\ correlation, we find
\begin{equation}
	\frac{\MBH}{10^8~\Msun}
	\sim
	0.11\left(\frac{\MDM}{10^{12}~\Msun}\right)^{1.27}.
\label{MBHMDM}
\end{equation}
It should be noted however, as indicated by Ferrarese~(2002), that the
uncertainties in this conversion can be quite large. For example, a
major uncertainty is how the baryonic infall in dark matter haloes
affects the baryonic rotation curve (e.g.~Seljak~2002). Therefore, the
correlation (\ref{MBHMDM}) shouls be regarded as a rough guideline
only. The \MBHvc\ correlation~(\ref{MBHvc}) on the other hand is based
solely on observed quantities, and has a much smaller
uncertainty. This correlation therefore serves as an important (purely
observational) constraint that should be reproduced and explained by
theoretical galaxy formation models. Combined with other tight
relations such as the \MBHsigma\ relation and the Tully-Fisher
relation, it clearly points at an intimate interplay between the
various components (dark matter, discs, bulges and SMBHs), and forms a
strong test for galaxy formation and evolution models.

Apart from being an ingredient in theoretical galaxy formation models,
the derived \MBHvc\ relation can also serve as a practical tool to
estimate the black hole content for spiral galaxies. Due to the large
scatter of the \MBHLB\ relation, the most reliable means of estimating
$\MBH$ in galaxies is by using the \MBHsigma\ relation. Unfortunately,
the number of spiral galaxies with reliable velocity dispersion
measurements is relatively small. On the contrary, extended rotation
curves have been measured for large samples of spiral galaxies, mainly
for use in Tully-Fisher relation studies (e.g.\ Persic \&
Salucci~1995; Courteau~1997; Verheijen \& Sancisi~2001). This makes
the tight \MBHvc\ correlation a practical tool to estimate the black
hole content of these galaxies. In particular, the \MBHvc\ correlation
can be used in a statistical way for SMBH demographics studies (e.g.\
Merritt \& Ferrarese~2001c; Ferrarese~2002b; Aller \& Richstone~2002;
Yu \& Tremaine~2002), by combining it with spiral galaxy velocity
functions (Shimasaku~1993; Gonzalez et al.~2000). This falls beyond
the scope of this Letter, and will be addressed in future work.

\end{document}